# Stochastic modeling of citation slips

M.V. Simkin and V.P. Roychowdhury
*Department of Electrical Engineering, University of California, Los Angeles, CA 90095-1594*

We present empirical data on frequency and pattern of misprints in citations to twelve high-profile papers. We find that the distribution of misprints, ranked by frequency of their repetition, follows Zipf's law. We propose a stochastic model of citation process, which explains these findings, and leads to the conclusion that about 70-90% of scientific citations are copied from the lists of references used in other papers.

## Introduction

Comparative studies of the popularity of scientific papers has been a subject of much research [1]-[8], but the scope has been limited to citation distributions. The present paper estimates the fraction of citations that were copied by the citing authors. A preliminary report of this study has been published before [9].
   Our study reveals that misprints in scientific citations are common and that many misprints are identical. The probability of repeating someone else's misprint accidentally is small. One concludes that repeat misprints are most likely due to copying from a reference list used in another paper. Our initial report [9] led to a lively public debate [10] on whether copying citations is tantamount to not reading the original paper. Alternative explanations are worth exploring; however, such hypotheses should be supported by data and not by anecdotal claims. It is indeed most natural to assume that a copying citer also failed to read the paper in question (albeit this can not be rigorously proved). *Entities must not be multiplied beyond necessity.* Having thus shaved the critique with Occam's razor, we will proceed to use the term non-reader to describe a citer who copies.   While several researchers [11]-[13] had noticed repeat misprints in citations and attributed them to copying references, no consistent attempt at estimating the fraction of copied reference in scientific literature was done until [9]. A bibliometrics textbook [5] lists thirteen reasons undermining the validity of the citation data, even coining new terms such as the "American factor", without any mention of citation copying. Thus, misprints in scientific citations have traditionally been treated as accidental: something to be merely corrected [14] but not analyzed (a fate similar to that of slips in speech and writing before Freud [15]).  We, in contrast, demonstrate that citation misprints provide a window into human dynamics driving the scientific citation process.

As misprints in citations are not too frequent, only celebrated papers provide enough statistics to work with. Figure 1 shows distributions of misprints (the caption of the figure provides the procedure used to detect misprints) in citations to four of such papers in the rank-frequency representation, introduced by Zipf [16].  Tables 1 and 2 summarize citation/misprints statistics for all 12 papers studied.

As a preliminary attempt, one can estimate the ratio of the number of readers (non-copiers) to the number of citers, $R$, as the ratio of the number of **distinct** misprints, $D$, to the **total number** of misprints, $T$. Clearly, we know  that among $T$ citers, $T - D$ copied, because they repeated someone else's misprint.  For the $D$ others, with the information at hand, we don't have any evidence that they copied, so according to the presumed innocent principle, we assume that they did write their own citation.  Then in our sample, we have $D$ readers and $T$ citers, which lead to:

$$R \approx D/T. \qquad (1)$$

The values of $R$ obtained using Eq.1 are given in Table 2 and range between 12% and 58% for different papers. The average is 27%. This estimate would be correct if the people who introduced original misprints had got the citation from the original paper. However, given the low values of $R$, it is obvious that many original misprints were introduced while copying references. Therefore, a more careful analysis is necessary. We need a model to accomplish it.

**Table 1.** Papers, misprints in citing which were studied.

| No | Reference |
|---|---|
| 1 | K. G. Wilson, Phys. Rev. 179, 1499 (1969) |
| 2 | K. G. Wilson, Phys. Rev. B 4, 3174 (1971) |
| 3 | K. G. Wilson, Phys. Rev. B 4, 3184 (1971) |
| 4 | K. G. Wilson, Phys. Rev. D 10, 2445 (1974) |
| 5 | J.M. Kosterlitz and D.J. Thouless, J. Phys. C 6, 1181 (1973) |
| 6 | J.M. Kosterlitz, J. Phys. C 7, 1046 (1974) |
| 7 | M.J. Feigenbaum, J. Stat. Phys. 19, 25 (1978) |
| 8 | M.J. Feigenbaum, J. Stat. Phys. 21, 669 (1979) |
| 9 | P. Bak, J. von Boehm, Phys. Rev. B 21, 5297 (1980) |
| 10 | P. Bak, C. Tang, and K. Wiesenfeld, Phys. Rev. Lett. 59, 381 (1987) |
| 11 | P. Bak, C. Tang, and K. Wiesenfeld, Phys. Rev. A 38, 364 (1988) |
| 12 | P. Bak and C. Tang, J. Geophys. Res. B 94, 15635 (1989) |

**Table 2.** Citation and misprint statistics together with estimates of $R$ for twelve studied papers.

| No. | Citations | Misprints | | M | R | | | | % rank for $R = 0.2$ |
| | | total | distinct | | Eq.1 | Eq.12 | Eq.14 | MC | |
|---|---|---|---|---|---|---|---|---|---|
| 1 | 1291 | 61 | 29 | 2.2% | 0.48 | 0.46 | 0.44 | 0.37 | 15% |
| 2 | 861 | 33 | 13 | 1.5% | 0.39 | 0.38 | 0.35 | 0.28 | 44% |
| 3 | 818 | 38 | 11 | 1.3% | 0.29 | 0.28 | 0.22 | 0.22 | 68% |
| 4 | 2578 | 263 | 32 | 1.2% | 0.12 | 0.11 | no solution | 0.10 | 95% |
| 5 | 4301 | 196 | 45 | 1.0% | 0.23 | 0.22 | 0.17 | 0.15 | 76% |
| 6 | 1673 | 40 | 12 | 0.7% | 0.30 | 0.29 | 0.25 | 0.22 | 65% |
| 7 | 1639 | 36 | 21 | 1.3% | 0.58 | 0.58 | 0.57 | 0.49 | 6% |
| 8 | 837 | 55 | 18 | 2.2% | 0.33 | 0.31 | 0.26 | 0.22 | 57% |
| 9 | 419 | 20 | 8 | 1.9% | 0.40 | 0.39 | 0.34 | 0.29 | 50% |
| 10 | 1717 | 33 | 14 | 0.8% | 0.42 | 0.42 | 0.40 | 0.31 | 36% |
| 11 | 1348 | 78 | 27 | 2.0% | 0.35 | 0.33 | 0.29 | 0.23 | 47% |
| 12 | 397 | 61 | 18 | 4.5% | 0.30 | 0.26 | 0.17 | 0.19 | 69% |

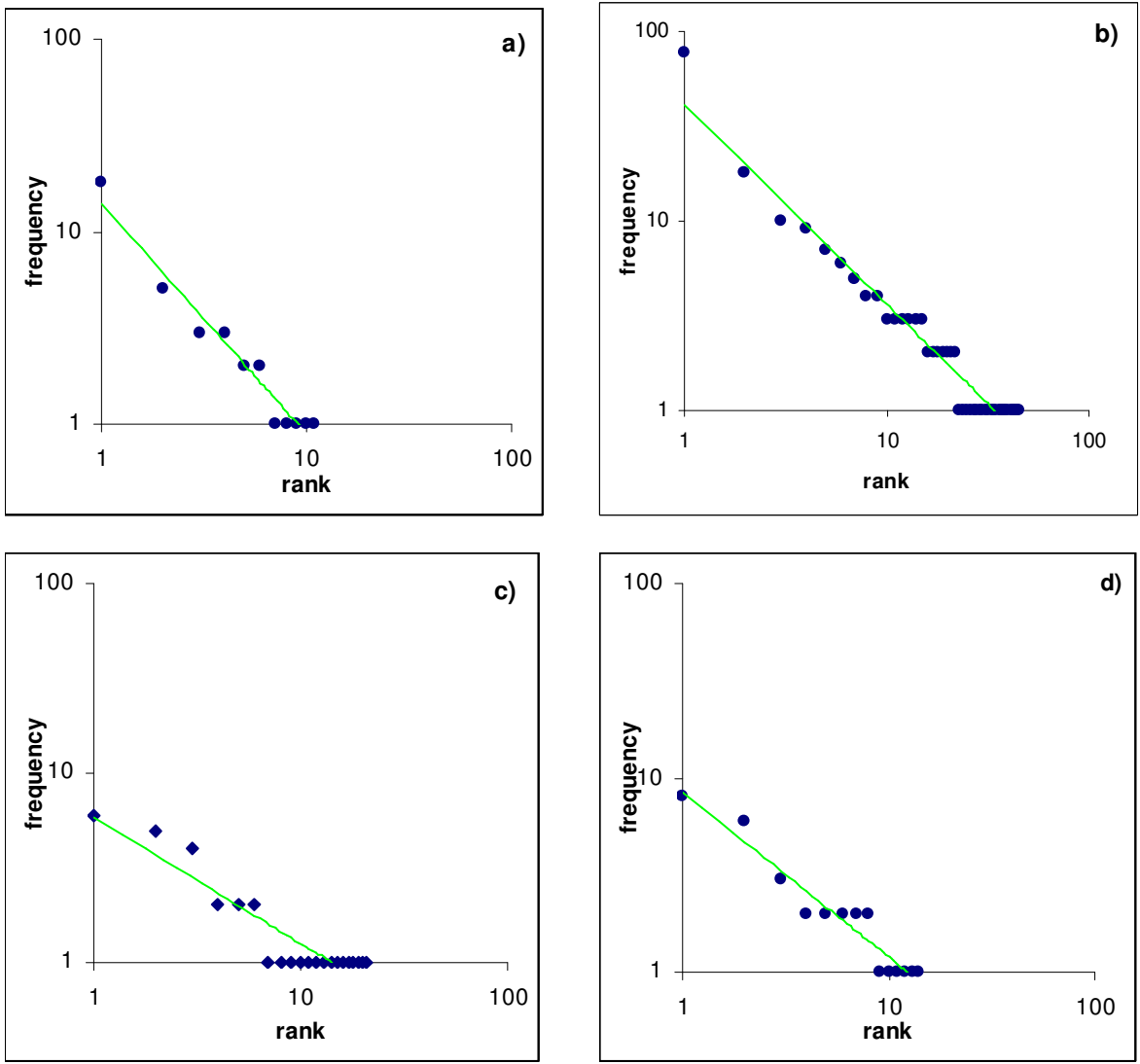

**Figure 1.** Rank-frequency distributions of misprints in referencing four high-profile papers. Figures a), b), c), and d) are for Papers 2, 5, 7, and 10 of Table 1. Solid lines are fits to Zipf Law with exponents a) 1.20; b) 1.05; c) 0.66; d) 0.85. The way we count misprints is look at the whole sequence of volume, page number and the year, which amounts to between 8 and 11 digits for different studied papers. That is, two misprints are distinct if they are different in any of the places, and they are repeats if they agree on *all* of the digits.

## Model for misprints propagation

Our misprints propagation model (MPM), which was stimulated by Simon's [17] explanation of Zipf Law and Krapivsky-Redner [18] idea of link redirection, is as follows. Each new citer finds the reference to the original in any of the papers that already cite it (or it can be the original paper itself). With probability $R$ he gets the citation information from the original. With probability 1-$R$ he copies the citation to the original from the paper he found the citation in. In either case, the citer introduces a new misprint with probability $M$.

Let us derive the evolution equations for the misprint distribution. The only way to increase the number of misprints that appeared only once, $N_1$, is to introduce a new misprint. So, with each new citation $N_1$ increases by one with probability $M$. The only way to decrease $N_1$, is to copy correctly one of misprints that appeared only once, this happens with probability $\alpha \times \dfrac{N_1}{N}$, where

$$\alpha = (1-R) \times (1-M) \qquad (2)$$

is the probability that a new citer copies the citation without introducing a new error, and $N$ is the total number of citations. For the expectation value we thus have:

$$\frac{dN_1}{dN} = M - \alpha \times \frac{N_1}{N}. \qquad (3a)$$

The number of misprints that appeared $K$ times, $N_K$, (where $K > 1$) can be increased only by copying correctly a misprint which appeared $K-1$ times. It can only be decreased by copying (again correctly) a misprint which appeared $K$ times. For the expectation values we thus have:

$$\frac{dN_K}{dN} = \alpha \times \frac{(K-1) \times N_{K-1} - K \times N_K}{N} \quad (K>1). \qquad (3b)$$

Assuming that the distribution of misprints has reached its stationary state we can replace the derivatives ($dN_K/dN$) by ratios ($N_K/N$) to get:

$$\frac{N_1}{N} = \frac{M}{1+\alpha},$$
$$\frac{N_{K+1}}{N_K} = \frac{K}{1+1/\alpha+K} \quad (K>1). \qquad (4)$$

Note that for large $K$: $N_{K+1} \approx N_K + dN_K/dK$, therefore Equation (4) can be rewritten as:

$$\frac{dN_K}{dK} \approx -\frac{1+1/\alpha}{1+1/\alpha+K} N_k \approx \frac{1+1/\alpha}{K} N_k.$$

From this follows that the misprints frequencies are distributed according to a power law:

$$N_K \sim 1/K^\gamma, \qquad (5)$$

where

$$\gamma = 1 + \frac{1}{\alpha} = 1 + \frac{1}{(1-R)\times(1-M)}. \qquad (6)$$

Relationship between $\gamma$ and $\alpha$ in Eq.(6) is the same as the one between exponents of number-frequency and rank-frequency distributions[1]. Therefore the parameter $\alpha$, which was defined in Eq.(2), turned out to be the Zipf law exponent.

An exact formula for $N_k$ can also be obtained by iteration of Eq.(4) to get:

$$\frac{N_K}{N} = \frac{\Gamma(K)\Gamma(\gamma)}{\Gamma(K+\gamma)} \times \frac{M}{\alpha} = B(K,\gamma) \times \frac{M}{\alpha} \qquad (7)$$

Here $\Gamma$ and $B$ are Euler's Gamma and Beta functions. Using the asymptotic for constant $\gamma$ and large $K$

$$\frac{\Gamma(\gamma)}{\Gamma(K+\gamma)} \sim K^{-\gamma} \qquad (8)$$

we recover Eq.(6).

The rate equation for the total number of misprints is:

$$\frac{dT}{dN} = M + \alpha \times \frac{T}{N}. \qquad (9)$$

The stationary solution of Eq. (9) is:

$$\frac{T}{N} = \frac{M}{1-\alpha} = \frac{M}{R+M-RM}. \qquad (10)$$

---

[1] Suppose that the number of occurrences of a misprint ($K$), as a function of the rank ($r$), when the rank is determined by the above frequency of occurrence, follows a Zipf law: $K(r) = \frac{C}{r^\alpha}$. We want to find the number-frequency distribution, i.e. how many misprints appeared $n$ times. The number of misprints that appeared between $K_1$ and $K_2$ times is obviously $r_2 - r_1$, where $K_1 = C/r_1^\alpha$ and $K_2 = C/r_2^\alpha$. Therefore, the number of misprints that appeared $K$ times, $N_k$, satisfies $N_K dK = -dr$ and hence, $N_K = -dr/dK \sim K^{-1/\alpha - 1}$.

The expectation value for the number of distinct misprints is obviously

$$D = N \times M. \quad (11)$$

From Equations (10) and (11) we obtain:

$$R = \frac{D}{T} \times \frac{N-T}{N-D}, \quad (12)$$

The values obtained using Eq.(12) are given in Table 1. They range between 11% and 58%.

One can ask why we did not choose to extract $R$ using Equations (2) or (6). This is because $\alpha$ and $\gamma$ are not very sensitive to $R$ when it is small (in fact Eq. 4 gives negative values of $R$ for some of the fittings in Figure 1). In contrast, $T$ scales as $1/R$.

We can slightly modify our model and assume that original misprints are only introduced when the reference is derived from the original paper, while those who copy references do not introduce new misprints (e.g. they do cut and paste). In this case one can show that $T = N \times M$ and $D = N \times M \times R$. As a consequence Eq.(1) becomes exact (in terms of expectation values, of course).

**Finite-size corrections**

Preceding analysis assumes that the stationary state had been reached. Is this reasonable? Eq.(9) can be rewritten as:

$$\frac{d(T/N)}{M - (T/N) \times (1-\alpha)} = d \ln N. \quad (13)$$

Naturally the first citation is correct (it is the paper itself). Then the initial condition is $N = 1; T = 0$. Eq.(13) can be solved to get:

$$\frac{T}{N} = \frac{M}{1-\alpha} \times \left(1 - \frac{1}{N^{1-\alpha}}\right) = \\ \frac{M}{R+M-M \times R} \times \left(1 - \frac{1}{N^{R+M-M \times R}}\right) \quad (14)$$

This should be solved numerically for $R$. The values obtained using Eq.(14) are given in Table 1. They range between 17% and 57%. Note that for one paper (No.4) no solution to Eq. (14) was found[2]. As $N$ is not a continuous variable, integration of Equation 9 is not perfectly justified, particularly when $N$ is small. Therefore we reexamine the problem using a rigorous discrete approach due to Krapivsky and Redner [19]. The total number of misprints, $T$, is a random variable that changes according to

$$T(N+1) = \begin{cases} T(N) & \text{prob. } 1 - M - \frac{T(N)}{N}\alpha \\ T(N)+1 & \text{prob. } M + \frac{T(N)}{N}\alpha \end{cases} \quad (15)$$

after each new citation. Therefore the expectation values of $T$ obey following recursion relations:

$$\langle T(N+1) \rangle = \langle T(N) \rangle + \frac{\langle T(N) \rangle}{N}\alpha + M \quad (16)$$

To solve Eq. (16) we define a generating function:

$$\chi(\omega) = \sum_{N=1}^{\infty} \langle T(N) \rangle \omega^{N-1} \quad (17)$$

After multiplying Eq. (16) by $N\omega^{N-1}$ and summing over $N \geq 1$ the recursion relation is converted into the differential equation for the generating function

$$(1-\omega)\frac{d\chi}{d\omega} = (1+\alpha)\chi + \frac{M}{(1-\omega)^2} \quad (18)$$

---

[2] Why did this happen? Obviously, $T$ reaches maximum when $R$ equals zero. Substituting $R = 0$ in Eq.(14) we get: $T_{MAX} = N \times (1 - 1/N^M)$. For paper No.4 we have $N = 2,578$, $M = D/N = 32/2,578$. Substituting this into the preceding equation we get $T_{MAX} = 239$. The observed value $T = 263$ is therefore higher than an expectation value of $T$ for any $R$. This does not immediately suggest discrepancy between the model and experiment but a strong fluctuation. In fact out of 1,000,000 runs of Monte-Carlo simulation of MPM with the parameters of the mentioned paper and $R=0.2$ exactly 49,712 runs (almost 5%) produced $T \geq 263$.

Solving Eq.(18) subject to the initial condition $\chi(0) = \langle T(1) \rangle = 0$ gives

$$\chi(\omega) = \frac{M}{1-\alpha}\left(\frac{1}{(1-\omega)^2} - \frac{1}{(1-\omega)^{1+\alpha}}\right) \quad (19)$$

Finally we expand the right hand side of Eq. (19) in Taylor series in $\omega$ and equating coefficients of $\omega^{N-1}$ obtain:

$$\frac{\langle T(N) \rangle}{N} = \frac{M}{1-\alpha}\left(1 - \frac{\Gamma(N+\alpha)}{\Gamma(1+\alpha)\Gamma(N+1)}\right) \quad (20)$$

Using Eq. (8) we obtain that for large $N$

$$\frac{\langle T(N) \rangle}{N} = \frac{M}{1-\alpha}\left(1 - \frac{1}{\Gamma(1+\alpha)} \times \frac{1}{N^{1-\alpha}}\right) \quad (21)$$

This is identical to Eq. (14) except for the pre-factor $1/\Gamma(1+\alpha)$. Parameter $\alpha$ (it is defined in Eq. (2)) ranges between 0 and 1. Therefore, the argument of Gamma-function ranges between 1 and 2. Because $\Gamma(1) = \Gamma(2) = 1$ and between 1 and 2 Gamma function has just one extremum $\Gamma(1.4616...) = 0.8856...$, the continuum approximation (Eq.(14)) is reasonably accurate.

## Monte-Carlo simulations

In the preceding section we calculated the expectation value of $T$. It does not, however, necessarily coincide with the most likely value when the probability distribution is not Gaussian. To get a better idea of the models behavior for small $N$ and a better estimate of $R$ we did numerical simulations. To simplify comparison with actual data the simulations were performed in a "micro-canonical ensemble", i.e. with a fixed number of distinct misprints. Each paper is characterized by the total number of citations, $N$, and the number of distinct misprints, $D$. At the beginning of a simulation $D$ misprints are randomly distributed between $N$ citations and chronological numbers of the citations with misprints are recorded in a list. In the next stage of the simulation for each new citation, instead of introducing a misprint with probability $M$, we introduce a misprint only if its chronological number is included in the list created at the outset. This way one can ensure that the number of distinct misprints in every run of a simulation is equal to the actual number of distinct misprints for the paper in question. A typical outcome of such simulation for Paper No.5 is shown in Fig. 2.

To estimate the value of $R$, 1,000,000 runs of the random-citing model with $R = 0, 0.1, 0.2... 0.9$ were done. An outcome of such simulations for one paper is shown in Fig. 3. Number of times, $N_R$, when the simulation produced a total number of misprints identical to the one actually observed for the paper in question was recorded for each $R$. Bayesian inference was used to estimate the probability of $R$:

$$P(R) = \frac{N_R}{\sum_R N_R} \quad (22)$$

Estimated probability distributions of $R$, computed using Eq. (22) for four sample papers are shown in Figure 4. The median values are given in Table 2 (see the MC column). They range between 10% and 49%.

Now let's assume $R$ to be the same for all twelve papers and compute Bayesian inference:

$$P(R) = \frac{\prod_{i=1}^{12} N_R^i}{\sum_R \prod_{i=1}^{12} N_R^i} \quad (23)$$

The result is shown in Fig. 5. $P(R)$ is sharply peaked around $R=0.2$. The median value of $R$ is 18%. And with 95% probability $R$ is less than 34%. This low value is consistent with the "Principle of Least Effort" [16].

But is the assumption that $R$ is the same for all twelve papers reasonable? (The estimates for separate papers vary between ten and fifty percent!) To answer this question the following analysis was done. Let us define for each paper a "percentile rank". This is the fraction of the simulations of the MPM (with $R = 0.2$) that produced $T$, which was less than actually observed $T$. Actual values of these percentile ranks for each paper are given in Table 2 and their cumulative distribution is shown in Fig. 6. Now if we claim that MPM (with same $R = 0.2$ for all papers) indeed describes the reality – then the distribution of these percentile ranks must be uniform. Whether or not the actual data is consistent with this can be checked using Kolmogorov-Smirnov test [20]. The maximum value of the absolute difference between two cumulative distribution functions (*D*-statistics) in our case is $D = 0.15$. According to the Kolmogorov-Smirnov test, the probability of *D* to be more than that is 91%. This means that the data is perfectly consistent with the assumption of $R = 0.2$ for all papers.

One can notice that the estimates of *M* (computed as $M=D/N$) for different papers (see Table 2) are also different. One may ask if it is possible that *M* is the same for all papers and different values of *D*/*N* are results of fluctuations. The answer is that the data is totally inconsistent with single *M* for all papers. This is not unexpected, because some references can be more error-prone, for example, because they are longer. Indeed, the most-misprinted paper (No.12) has two-digit volume number and five-digit page number.

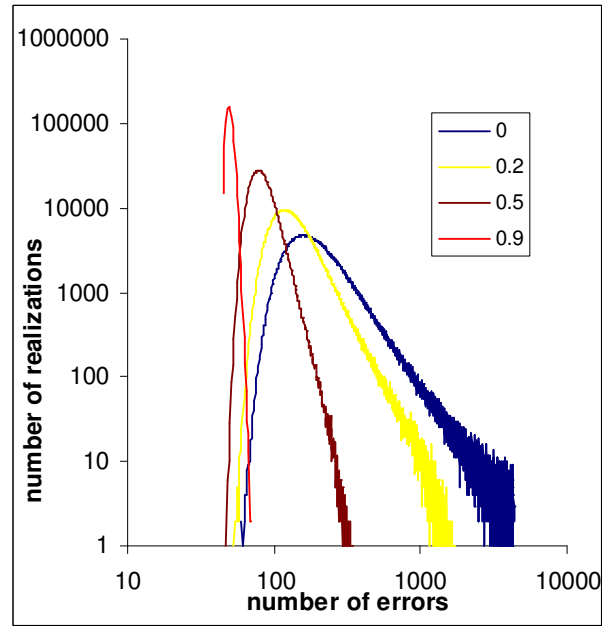

**Figure 3.** The outcome of 1,000,000 runs of the MPM with $N=4301$, $D=45$ (parameters of paper No.5 from Table 1) for four different values of *R*

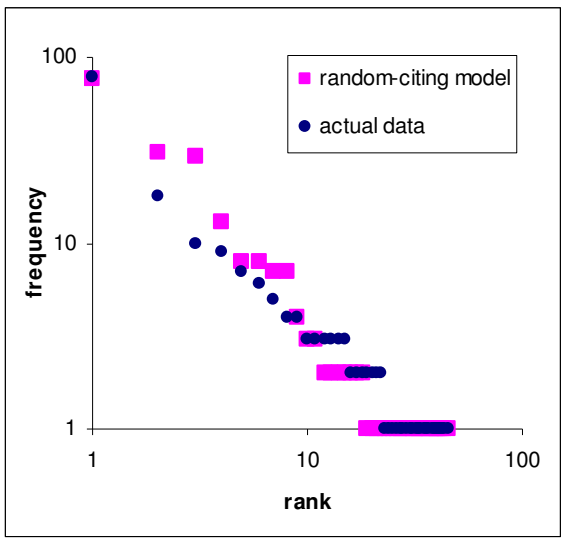

**Figure 2.** A typical outcome of a single simulation of the MPM (with $R=0.2$) compared to the actual data for Paper No.5 (Table 1).

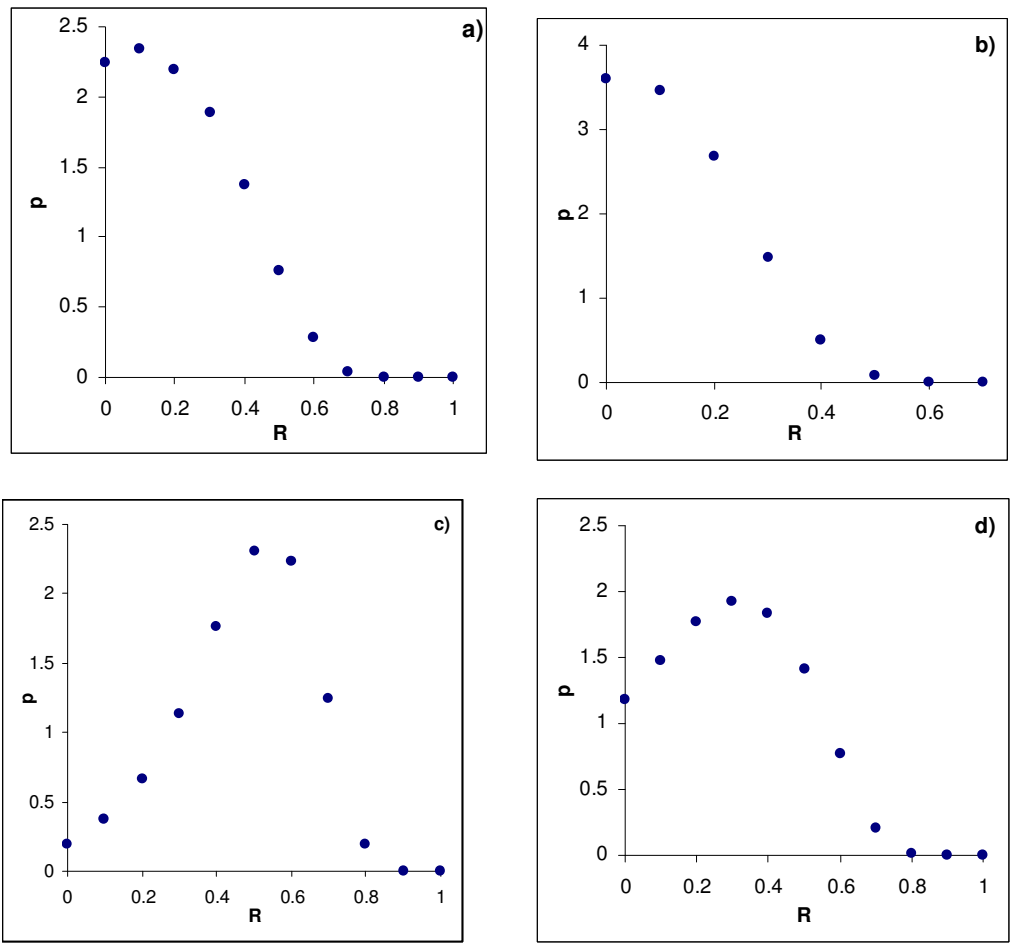

**Figure 4.** Bayesian inference for the readers/citers ratio, R, computed using Eq. (22). Figures a), b), c), and d) are for Papers No. 2, 5, 7, and 10 (Table 1).

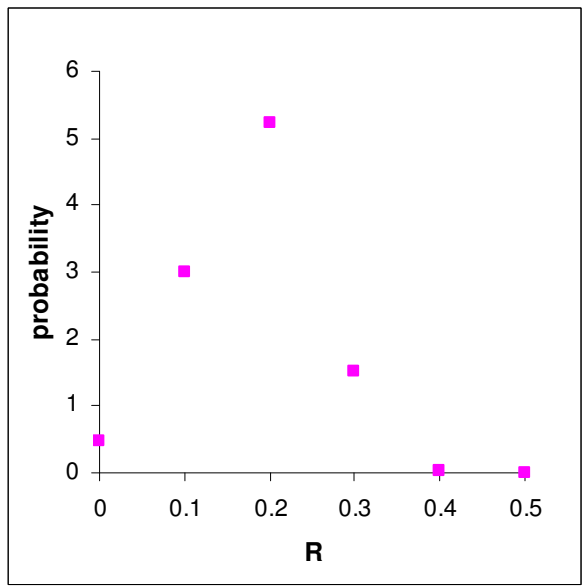

**Figure 5.** Bayesian inference for the readers/citers ratio, *R*, based on twelve studied papers computed using Eq.(23).

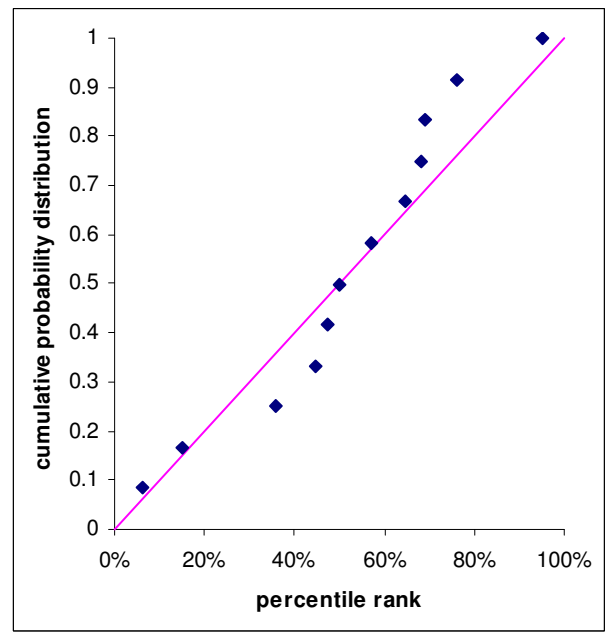

**Figure 6.** Cumulative distribution of the percentile ranks of the observed values of *T* with regard to the outcomes of the simulations of the MPM with *R* = 0.2 (diamonds). For comparison the cumulative function of the uniform distribution is given (a line).

## Operational limitations of the model

Our analysis is not perfect. There are occasional repeat identical misprints in papers, which share individuals in their author lists. To estimate the magnitude of this effect we took a close look at all 196 misprinted citations to paper No. 5 of Table 1. It turned out that such events constitute a minority of repeat misprints. It is not obvious what to do with such cases when the author lists are not identical: should the set of citations be counted as a single occurrence (under the premise that the common co-author is the only source of the misprint) or as multiple repetitions. Counting all of such repetitions as only a single misprint occurrence results in elimination of 39 repeat misprints. The number of total misprints, $T$, drops from 196 to 157, bringing the upper bound for $R$ (Eq.1) from $\frac{45}{196} \cong 23\%$ up to $\frac{45}{157} \cong 29\%$. An alternative approach is to subtract all the repetitions of each misprint by the originators of that misprint from non-readers and add it to the number of readers. There were 11 such repetitions, which increases $D$ from 45 up to 56 and the upper bound for $R$ (Eq.1) rises to $\frac{56}{196} \cong 29\%$, which is the same value as the preceding estimate. It would be desirable to redo the estimate using Equations 12 and 14, but the misprint propagation model would have to be modified to account for repeat citations by same author and multiple authorships of a paper. This may be a subject of future investigations.

Another issue brought up by the critics (e.g., [21]) is that because some misprints are more likely than others, it is possible to repeat someone else's misprint purely by chance. By examining the actual data one finds that about two third of distinct misprints fall in to the following categories:
  a) One misprinted digit in volume, page number, or in the year.
  b) One missing or added digit in volume or page number.
  c) Two adjacent digits in a page number are interchanged.

The majority of the remaining misprints are combinations of a), b), c) (e.g. one digit in page number omitted and one digit in year misprinted)[3]. For a typical reference there are over fifty aforementioned likely misprints. However, even if probability of certain misprint is not negligibly small but one in fifty, our analysis still applies. For example for paper No.5 (Table 1) the most popular error appeared 78 times, while there were 196 misprints total. Therefore, if probability of certain misprint is 1/50, there should be about $196/50 \approx 4$ such misprints, not 78. In order to explain repeat misprints distribution by higher probability of certain misprint this probability should be as big as $78/196 \approx 0.4$. This seems extremely unlikely. However, finding relative propensities of different misprints deserves further investigation.

Smith noticed [22] that some misprints are in fact introduced by the ISI. To estimate the importance of this effect we explicitly verified 88 misprinted (according to ISI) citations in the original articles. 72 of them were exactly as in the ISI database, but 16 where in fact correct citations. (To be precise some of them had minor inaccuracies, like second initial of the author was missing, while page number, volume and year where correct. Perhaps, they where victims of an "erroneous error correction" [22]). It is not clear how to consistently take into account these effects, specifically because there is no way to estimate how many wrong citations have been correctly corrected by ISI [14]. But given the relatively small percentage of the discrepancy between ISI database and actual articles ($16/88 \cong 18\%$) this can be taken as a noise with which we can live.

It is important to note that within the framework of the MPM $R$ is not the ratio of readers to citers, but the probability that a citer consults the original paper, provided that he encountered it through another paper's reference list. However, he could encounter the paper directly. This has negligible effect for highly-cited papers, but is important for low-cited papers. Within the MPM framework the probability of such an event for each new citation is obviously $1/n$, where $n$ is the current total number of citations. The expectation value of the true ratio of readers to citers is therefore:

$$R^*(N) = R + (1-R) \times \frac{\sum_{n=1}^{N} \frac{1}{n}}{N} \approx R + (1-R) \times \frac{\ln(2N+1)}{N}. \quad (24)$$

---
[3] There are also misprints where author, journal, volume and year are perfectly correct, but the page number is totally different. Probably, in such case the citer mistakenly took the page number from a neighboring paper in the reference list he was lifting the citation from.

The values of $R^*$ for papers with different total numbers of citations, computed using Eq.(24), are shown in Fig.7. For example, on average, about four people have read a paper which was cited ten times. One can use Eq.(24) and empirical citation distribution to estimate an average value of $R^*$ for the scientific literature in general. The formula is:

$$\langle R^* \rangle = \frac{\sum R^*(N_i) \times N_i}{\sum N_i} \quad (25)$$

Here the summation is over all of the papers in the sample and $N_i$ is the number of citations that $i$'s paper had received. The estimate, computed using citation data for Physical Review D [23] and Eqs (24) and (25) (assuming $R = 0.2$), is $\langle R^* \rangle \approx 0.33$.

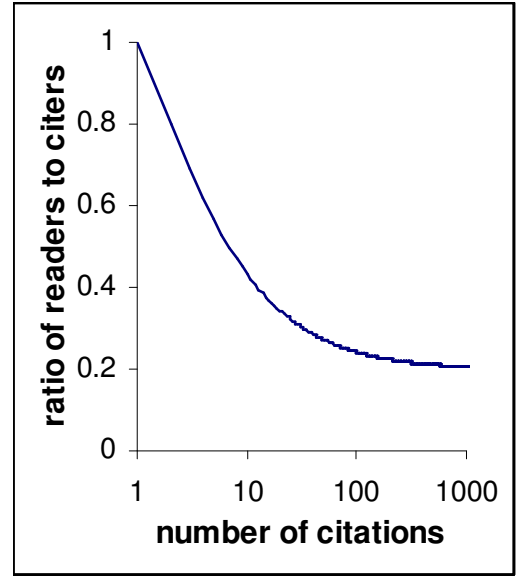

**Figure 7.** Ratio of readers to citers as a function of total amount of citations for $R = 0.2$, computed using Eq.(24).

## Relation to previous work

As it was mentioned before the bulk of previous literature on citations was concerned with their counting. After extensive literature search we found only a handful of papers which analyzed misprints in citations[4]. Broadus [11] looked through 148 papers which cited both the renowned book, which misquoted the title of one of its references, and that paper, the title of which was misquoted in the book. He found that 34 or 23% of citing papers made the same error as was in the book. Moed and Vries [12] (apparently independent of Broadus, as they don't refer to his work), found identical misprints in scientific citations and attributed them to citation copying. Hoerman and Nowicke [13] looked through a number of papers which deal with the so called Ortega Hypothesis of Cole and Cole [2]. When Cole and Cole quoted a passage from the book by

---

[4] A paper by Steel [24], titled identically to our first paper on misprints, i.e. "Read before you cite", turned out to deal with the subject using the analysis of the *content* of the papers, not of the propagation of misprints in references.

Ortega they introduced three distortions. Hoerman and Nowicke found seven papers which cite Cole and Cole and also quote that passage from Ortega. In six out of these seven papers all of the distortions made by Cole and Cole were repeated. According to [13] in this process even the original meaning of the quotation was altered[5].

While the fraction of copied citations found by Hoerman and Nowicke [13], $6/7 \cong 86\%$ agrees with our estimate, Boadus' number, 23%, seems to disagree with it. Note, however, that Broadus [11] assumes that citation, if copied - was copied from the book (because the book was renowned). Our analysis indicates that majority of citations to renowned papers are copied. Similarly, we surmise, in the Broadus' case citations to both the book and the paper were often copied from a third source.

## Copied citations - the mechanism of cumulative advantage

Almost forty years ago Merton observed [26] that the reward system in science is such that success is rewarded with an increased chance of further success. He called this Matthew effect (because "…unto every one that hath shall be given…" [Matthew 25:29])[6]. Later Price [3], building on the work of Simon [17], came up with a mathematical theory of Mathew Effect, which he called a cumulative advantage process. As far as bibliometry is concerned, the underlying model assumed that the rate of citation for a particular paper is proportional to the number of citations the paper had already received. This model was able to explain the previously observed power law in citation distribution [1]. However, no mechanism explaining why citation rate can be proportional to current amount of citations was proposed by Price.

More recently, cumulative advantage idea got a new life (and a new name) in the context of the world-wide web and other networks, where a power law distribution of connectivity was observed [27]. Kleinberg *et al* [28] suggested that such distribution can be achieved by simple link copying[7].
Several papers [18], [29], [30] have solved a number of related copying models and they indeed lead to a power law distribution of connectivity. *However, no evidence that such copying exists was provided.* Our analysis of misprint propagation *provides the evidence that citation copying dominates the dynamics of the network of scientific papers*. Recently we demonstrated [31] that a simple model, according to which when a scientist is writing a manuscript he picks three random[8] papers, cites them, and also copies a quarter of their references (also at random) accounts *quantitatively* for empirically observed distribution of citations.

The statement that a large amount of citations can be a result of a stochastic process, rather than a consequence of intrinsic merit may be shocking to some people. An indirect support for plausibility of such a conclusion comes from the very misprints frequencies – they follow the same Zipf law as the citation frequencies. Obviously no misprint is more seminal than the other[9].

---

[5] In fact information is sometimes *defined* by its property to deteriorate in chains [25].

[6] In fact similar saying appears in two other gospels: "For he that hath, to him shall be given…" [Mark 4:25], "…unto every one which hath shall be given…" [Luke 19:26] and belong to Jesus. Nonetheless the name "Mathew Effect" was repeated by hundreds of people who did not read the Bible.

[7] Another interesting idea is that citation dynamics is a self-organized critical process [32]

[8] Random-citing model is used not to ridicule the scientists, but because it can be exactly solved using available mathematical methods.

[9] A noteworthy case where prominence is reached by pure chance is the statistics of baby-names. It was observed that their frequency distribution follows a power law, and a copying mechanism that can explain this observation was proposed [33]. For example, during the 1990s 21,243 new-born American babies were named Michael, while only 55 were named Samson [34]. This means

We conclude with an excruciatingly apt citation (first used in this respect by Turner and Chubin [35] in their polemic with Cole and Cole [2]):

"…the race is not to the swift, nor the battle to the strong, neither yet bread to the wise, nor yet riches to men of understanding, nor yet favour to men of skill; but time and chance happeneth to them all." [Ecclesiastes 9:11].

---

that name "Michael" is almost four hundred times more popular then name "Samson". Is it intrinsically better?